# Signal Optimization with HV divider of MCP-PMT for JUNO

Fengjiao Luo[1,2,3], Zhimin Wang[2], Zhonghua Qin[1,2], Anbo Yang[1,2,3], Yuekun Heng[1,2]

[1] State Key Laboratory of Particle Detection and Electronics
(Institute of High Energy Phys-ics, CAS)
[2] Institute of High Energy Physics, CAS
[3] University of Chinese Academy of Science
`wangzhm@ihep.ac.cn`

On behalf of the JUNO Collaboration

**Abstract.** The Jiangmen Underground Neutrino Observatory (JUNO) is proposed to determine the neutrino mass hierarchy using a 20 kiloton underground liquid scintillator detector (CD). One of the keys is the energy resolution of the CD to reach <3% at 1 MeV, where totally 15,000 MCP-PMT will be used. The optimization of the 20" MCP-PMT is very important for better detection efficiency and stable performance. In this work, we will show the study to optimize the MCP-PMT working configuration for charge measurement. Particularly, the quality of PMT signal is another key for high-precision neutrino experiments while most of these experiments are affected by the overshoot of PMT signal from the positive HV scheme. The overshoot coupled with positive HV which is troubling trigger, dead time and precise charge measurement, we have studied to control it to less than ~1% of signal amplitude for a better physics measurement. In this article, on the one hand, the optimized HV divider ratio will be presented here to improve its collection efficiency; on the other hand, we will introduce the method to reduce the ratio of overshoot from 10% to 1%.

**Keywords:** JUNO, HV divider, MCP-PMT, collection efficiency, overshoot.

## 1  Introduction

The Jiangmen Underground Neutrino Observatory (JUNO) [1] will be located 53 km away from both Yangjiang and Taishan Nuclear Power Plants at Kaiping of Jiangmen of Guangdong province, in China. The JUNO project is proposed to determine the neutrino mass hierarchy using a 20 kton underground liquid scintillator detector (CD). As a multipurpose underground neutrino observatory, JUNO will also measure the neutrino oscillation parameters to better than 1% accuracy and measure neutrinos or antineutrinos from terrestrial and extra-terrestrial sources. To achieve ~3% at 1MeV, JUNO detector is designed with ~75% photo-cathode coverage; in total 15000 20'' MCP-PMTs from North Night Vision Technology Co. (NNVT) and 5000 20'' Dynode-PMTs from Hamamatsu will be used. High detection efficiency of PMT is one key of JUNO project, and detection efficiency is equal to quantum efficiency multiply by collection efficiency, it is essential to optimize the HV divider of PMT to reach maximum collection efficiency of PMT as shown in section 2.

The positive HV scheme is used on these PMTs and overshoot following a signal will be unavoidable. The overshoot will cause problems for precise charge measurement, system triggering and dead time as shown by Double Chooz [2], KamLAND [3], SNO [4], Borexino [5] and Daya Bay [6] experiments. In this paper, we will take measures to reduce the overshoot form 10% to 1% and make clear the reason of overshoot in section 3.

## 2 Optimization of the collection efficiency with HV divider

High photon detection efficiency (PDE) of PMT is one of a key for JUNO project, and PDE is the product of the photocathode quantum efficiency (QE) and the photoelectron collection efficiency (CE). The CE is designed by the photocathode shape, the shape and arrangement of the focusing electrode and supply voltage [7]. For dynode PMT from Hamamatsu, we get the optimal HV divider ratio from its company, and in this section, we mainly aimed to get the optimal configuration of HV divider ratio to realize its maximum collection efficiency of MCP-PMT. The measurement of absolute CE of PMT seemed to be impossible. And in this section, we intend to measure its relative PDE to achieve its maximum CE by optimizing its HV divider ratio between photocathode (K) and focus electrode (F1 and F2). The circuit of HV divider for MCP-PMT is shown as Fig.1. The distribution of R1 and R2 are the key of our study keeping the total HV divider ratio the same.

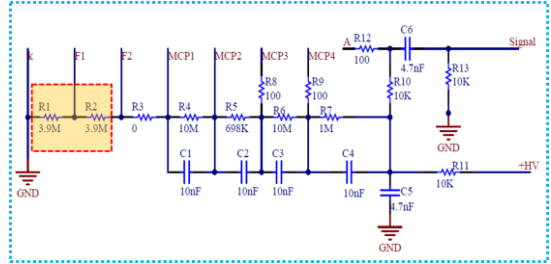

**Fig. 1.** Circuit of HV divider of MCP-PMT

### 2.1 Experiment setup

PDE is defined as the ratio of the number of counted pulses (output pules) to the number of incident photons. In this section, to obtain its optimal configuration of HV divider, we compared its relative PDE of MCP-PMT under different configurations. When keep the input photons the same, relative PDE can be obtained by the output pulses of PMT. The number of output pulses larger, the relative PDE better. The output pulses can be defined as the output charge divided by charge of single photoelectron. The difficult of the experiment is to keep the input light intensity the same under different HV divider ratio. A 20-inch Hamamatsu dynode PMT R12860 is adopted to be a reference to solve the question.

The experiment setup scheme is shown in Fig.2. The tested MCP-PMT and reference Hamamatsu PMT are all put in the dark box with EMF shielding. The light source is LED (+ diffuse box) with wavelength at 410 nm which is driven by a program controlled pulse generator. The high voltage was provided by a CAEN Mod.A1733P in CAEN SY4527 crate, and it can provide the high voltage for 2 PMTs respectively. The PMT anode signal of the 2 PMTs are sent to a FADC DT5751 to sample the waveform at the same time. Before PMT measurement, PMTs are preheated 2 hours with normal HV at gain of $10^7$ in dark room to keep its dark noise stable.

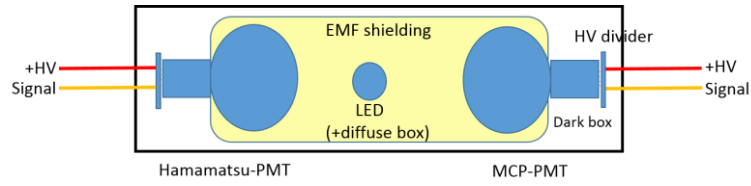

**Fig.2.** Experiment setup scheme

## 2.2 Optimization of the HV divider ratio

After the preheated of PMTs , at first, we obtain the single photoelectron (SPE) spectrum at the gain of $1*10^7$ of 2 PMTs as shown in Fig.3. Under the gain at $1*10^7$, the charge spectrum of 2 PMTs are measured at the same time varying the light intensity. The mean number of output photoelectrons are acquired. And we can obtain the relationship of relative detection efficiency between MCP-PMT and Hamamatsu PMT as shown in Fig.4 (a). The slope of the fitting line can be measured as the relative PDE of MCP-PMT.

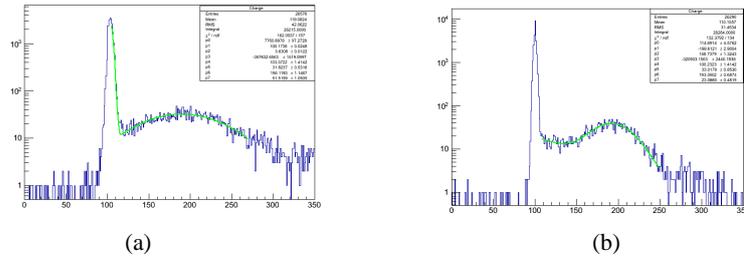

(a)                                     (b)

**Fig.3.** SPE spectrum of MCP-PMT (a) and Hamamatsu PMT (b) under the gain of $1*10^7$.

Each measurement of different HV divider ratio, only exchanged the HV divider of MCP-PMT and keep the experiments conditions the same, HV divider ratio as shown Table 1. The relative slope of MCP-PMT as a function of the ratio of (K-F1) / (F1-F2) are shown as Fig.4 (b).

**Table 1.** HV divider ratio

| K-F1:F1-F2 | 1.2:6.8 | 2.4:5.6 | 3.3:4.7 | 3.9:3.9 | 4.7:3.3 | 5.1:2.9 | 5.6:2.4 | 6.1:1.9 | 6.8:1.2 |
|---|---|---|---|---|---|---|---|---|---|
| (K-F1)/(F1-F2) | 0.176 | 0.429 | 0.892 | 1.000 | 1.424 | 1.759 | 2.333 | 3.211 | 5.667 |

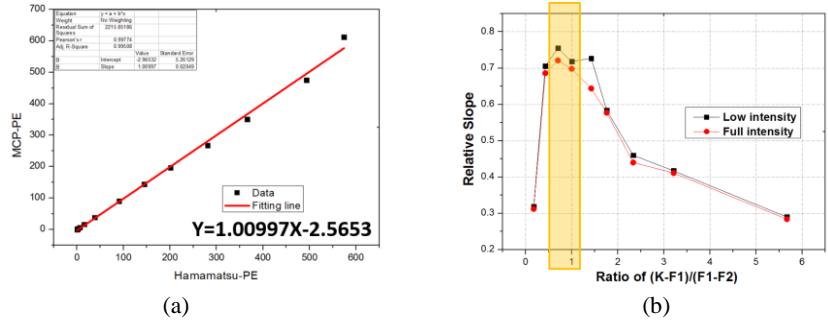

(a)                                                      (b)

**Fig.4.** Optimization of the HV divider ratio of MCP-PMT: (a) the relative DE of MCP-PMT; (b) the relative slope versus the ratio of (K-F1) / (F1-F2), the red line is the relative slope of PMT with full intensity from 0-600 PE, and the black line is the PMT with low intensity from 0-10 PE.

From Fig.4 (b), the relative slope will appear a peak at the ratio of (K-F1) / (F1-F2) around at 1. The electric field of MCP-PMT is optimal when the supplied voltage between K-F1 and F1-F2 is equal, and the photoelectrons can be pulled into the microchannel plate.

### 2.3    Results

The optimization of the working configuration of MCP-PMT to reach its optimal CE has been finished. We measured the relative DE to optimize its HV divider ratio between K-F1 and F1-F2. When the voltage between K-F1 and F1-F2 are the same high voltage, the relative DE will be the maximum. When the high voltage K-F1 and F1-F2 are differ greatly, the DE will decrease sharply. The electric field will has a big influence on CE. Combined the design and manufacture of MCP-PMT, the high voltage between K-F1 and F1-F2 should be almost the same to reach its maximum CE.

## 3    Optimization of signal overshoot

The quality of PMT signal is a key for large-size and high-precision neutrino experiments. PMT anode signal followed by an overshoot will affect its precision charge measurement, dead time and triggering time. In the following section, we study the mechanism of overshoot and successfully reduce the ratio of overshoot to less than 1% of signal amplitude for better physics measurement. A simplified HV divider with positive scheme is shown in Fig.5.

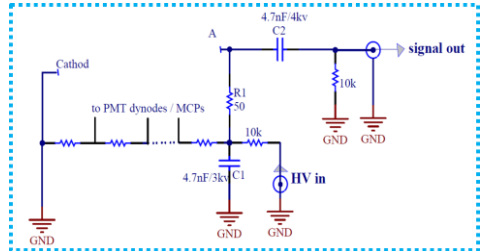

**Fig.5.** Simplified of the HV divider with positive scheme

### 3.1 Overshoot with 10%

With the positive scheme, we measured a large overshoot as shown in Fig.6 (a). The big overshoot originates from the decoupling capacitors in HV divider. The charge is collected by the PMT anode and the charging and discharging of capacitors C1 and C2 will affect the overshoot of anode signal. In Fig.6 (a), the green is our output signal from PMT. The output is big signal under intensive light by LED. LED is driven by the pulse generator and at the same time, a synchronized gate (blue line) from the pulse generator was sent to a low discrimination (LTD) to trigger oscilloscope.

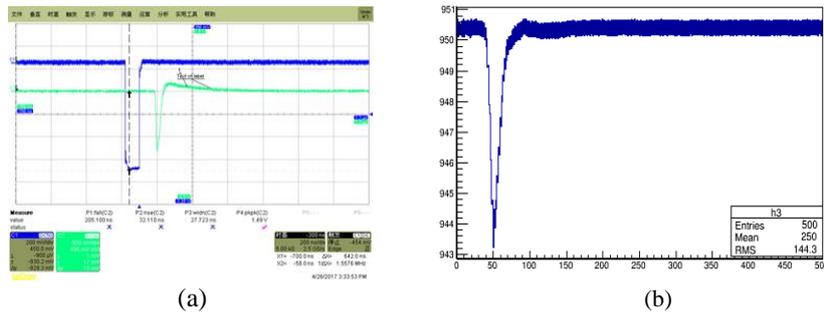

(a)             (b)

**Fig.6.** (a) Output signal of PMT with large overshoot, 500 mV/div, and 200 ns/div, the signal amplitude is about 1.25 V, overshoot is clear and the amplitude is about 200 mV or 16%; (b) Output signal of PMT with lower overshoot less than 1%.

### 3.2 Optimization of overshoot

The discharge of C1*R1 and C2*R2 is the consequence of overshoot, in our experiments, we take R1 from 50 ohm to 10 kohm by extended the discharging time, and the output signal as shown in Fig.6(b), the overshoot is less than 1% . In article [8] the study of the overshoot mechanism and model are reported.

## 4    Conclusion

In this work, we confirmed that the HV divider ratio between photocathode and focus electrode will have effect on the detection efficiency of MCP-PMT and designed an experimental method to measure its relative detection efficiency. The electric field of the type MCP-PMT will be at the optimal configuration when the voltage between K-F1 and F1-F2 is almost the same. Further optimization is still needed for the future JUNO HV divider to meet other detailed requirements. Besides, we have successfully controlled the ratio of overshoot to signal amplitude less than 1%, a smaller overshoot for a positive voltage PMT is important for charge measurement in high precision neutrino experiment.

## References

1. An F et al, Neutrino Physics with JUNO, J. Phys. G, 36:030401 (2016).
2. Y. Abe et al, Phys. Rev. D, 86 (5): 052008 (2012).
3. S. Abe et al, Phys. Rev. Lett, 100 (22): 221803 (2008).
4. http://www.sno.phy.queensu.ca/sno/str/SNO-STR-95-010.pdf, retrieved fourth June 2017.
5. J. Q. Meindl, Reconstruction and Measurement of Cosmogenic Signals in the Neutrino Experiment Borexino (Ph.D Thesis), Munchen: Technische Universitat Munchen, 2013.
6. F. Beissela, A. Cabrera, A. Cucoanes et al, Journal of Instrumentation, 8: T01003 (2013).
7. Hamamatsu Photonics K. K, Photomultiplier Tubes: Basics and Applications, Third edition (2007).
8. F. J. Luo, Y. K. Heng, Z. M. Wang et al, Chin. Phys. C, 40 (9): 481-486 (2017).